\documentclass[
    ,final            
  ]
  {aipproc}

\layoutstyle{8x11single}

\begin{document}

\title{Supernova Remnants and GLAST}


\author{Patrick Slane}{
  address={Harvard-Smithsonian Center for Astrophysics}
}

%

\begin{abstract}
It has long been speculated that supernova remnants represent a major
source of cosmic rays in the Galaxy. Observations over the past decade
have ceremoniously unveiled direct evidence of particle acceleration in
SNRs to energies approaching the knee of the cosmic ray spectrum.
Nonthermal X-ray emission from shell-type SNRs reveals multi-TeV
electrons, and the dynamical properties of several SNRs point to
efficient acceleration of ions. Observations of TeV gamma-ray emission
have confirmed the presence of energetic particles in several remnants
as well, but there remains considerable debate as to whether this
emission originates with high energy electrons or ions. Equally
uncertain are the exact conditions that lead to efficient particle
acceleration.

Based on the catalog of EGRET sources, we know that there is a large
population of Galactic gamma-ray sources whose distribution is similar
to that of SNRs. With the increased resolution and sensitivity of
GLAST, the gamma-ray SNRs from this population will be identified.
Their detailed emission structure, along with their spectra, will
provide the link between their environments and their spectra in other
wavebands to constrain emission models and to potentially identify
direct evidence of ion acceleration in SNRs. Here I 
summarize recent observational and theoretical work in the area of
cosmic ray acceleration by SNRs, and discuss the contributions GLAST
will bring to our understanding of this problem.
\end{abstract}

\maketitle


\section{Introduction}

Supernova remnants represent an energetic class of objects that is
intimately connected to sources of $\gamma$-ray emission in our
Galaxy. Shock acceleration by the SNR blast wave provides ample
energy for the production of multi-TeV particles, and the presence
of nearby material in dense clouds from which the remnant progenitors
collapsed forms a natural target for pion production with subsequent
$\pi^0 \rightarrow \gamma \gamma$ decay.  Nonthermal bremsstrahlung
of electrons off ambient material, as well as inverse Compton (IC)
scattering of electrons off ambient photons, can also lead to
$\gamma$-ray production -- and in many cases are the dominant
mechanisms.  SNRs are thus strong candidates for the emission of
$\gamma$-rays. X-ray observations have revealed numerous remnants
for which nonthermal emission from their shells provides direct
evidence of TeV electrons.  More recently, TeV $\gamma$-ray emission
has been detected from a number of SNRs, possibly providing evidence
for acceleration of ions as well.

The EGRET source catalog contains a large number of unidentified
sources in the Galactic plane, some of which have been tentatively
associated with SNRs, and such associations are plausible -- at
least in principle; particularly for remnants that are expanding
into dense environments, we expect emission that extends well into
the EGRET band. However, the large positional uncertainties for
these sources, along with the large sizes and high number density
of SNRS, yield a relatively high probability for false identifications
of EGRET sources with SNRs. Yet the emission in this band holds
particular importance in helping to understand the emission at
higher energies, and in addressing the overall picture of cosmic-ray
acceleration in SNRs. Below I discuss the production of $\gamma$-ray
emission from SNRs and describe recent results on evidence for
cosmic-ray acceleration in SNRs, with particular emphasis on
contributions that are expected from GLAST observations.

\section{Particle Acceleration by SNR Shocks}

As the blast wave from a supernova explosion expands, the surrounding
circumstellar material (CSM) and interstellar medium (ISM) is swept
up into a shell of hot gas. The shock jump conditions yield an
increase in density by a factor of $n/n_0 = (\gamma +1)/(\gamma -1)$, and
the associated postshock temperature is given by
\begin{equation}
T = \frac{2 (\gamma -1)}{(\gamma + 1)^2} \frac{\mu}{k} m_p V_s^2
\end{equation}
where $\gamma$ is the adiabatic index for the gas ($\gamma = 5/3$
for an ideal gas), $V_s$ is the shock speed, $m_p$ is the proton
mass, and $\mu$ is the mean molecular weight ($\mu \approx
0.6$ for solar abundances).  This shock-heated gas produces X-ray
emission characterized by a thermal bremsstrahlung spectrum accompanied
by emission lines.
As the blast wave decelerates upon sweeping up increasing amounts
of material, a reverse shock propagates back into the ejecta.  At
early times, the X-ray spectrum is dominated by emission from the
reverse-shocked ejecta; as the amount of swept-up material increases,
the spectrum becomes dominated by emission from material with ISM
abundances.

In addition to thermal heating of the swept-up gas, some fraction
of the shock energy density goes into production of relativistic
particles through diffusive shock acceleration wherein energetic
particles streaming away from the shock form turbulent waves which
act to scatter other particles back toward the shock. Subsequent
reacceleration builds up a nonthermal population of high energy
particles, with the maximum energy being limited by radiative
losses, the age of the SNR, or particle escape.  Such particle
acceleration by SNR shocks has long been suggested as a process by
which cosmic rays are produced, and radio observations provide
direct evidence of nonthermal electrons with energies of several
GeV.

If the relativistic particle component of the energy density becomes
comparable to that of the thermal component, the shock acceleration
process can become highly nonlinear. The gas becomes more compressible
(with $\gamma$ becoming much larger than 5/3), which results in a
higher density and enhanced acceleration.  The resulting efficient
production of nonthermal particles has a significant impact on the
dynamical evolution of the shock. Figure 1 (left, from \cite{Ell07})
shows the SNR temperature as a function of radius at
an age of 500 yr, assuming expansion into a uniform density $n_0 =
0.1 {\rm\ cm}^{-3}$ with an ambient magnetic field strength
$B_0 = 15 \mu$G. The short-dashed curve (labeled TP) corresponds
to the test-particle case in which no diffusive shock acceleration
operates.  The regions corresponding to the forward shock (FS),
the reverse shock (RS) and the contact discontinuity (CD) that
separates the swept-up and ejecta components are indicated.
Also shown are temperature plots for cases of moderate ($\epsilon
= 36\%$) and efficient ($\epsilon = 63\%$) particle acceleration.
(Here $\epsilon$ refers to the fraction of the energy flux crossing
the shock that ends up in relativistic particles.)
Two distinct observable effects are immediately evident: at a given
age, the separation between the FS and the RS (or CD) decreases
considerably with increased particle acceleration; and the temperature
at the forward shock is reduced for cases of high acceleration
efficiency.

\begin{figure}[t]
  \centerline{
  \includegraphics[height=.3\textheight]{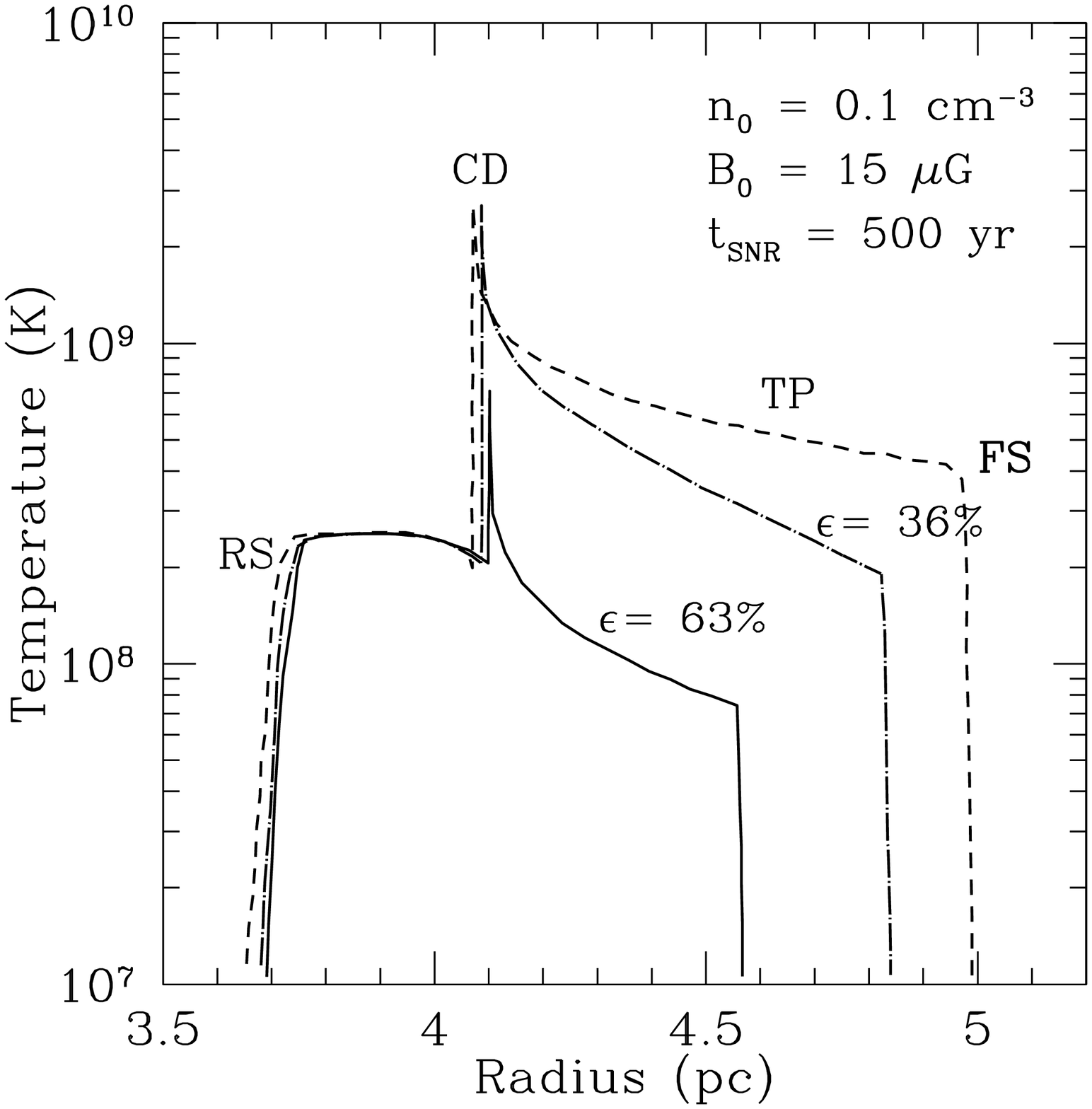}
  \hspace{0.3in}
  \includegraphics[height=.28\textheight]{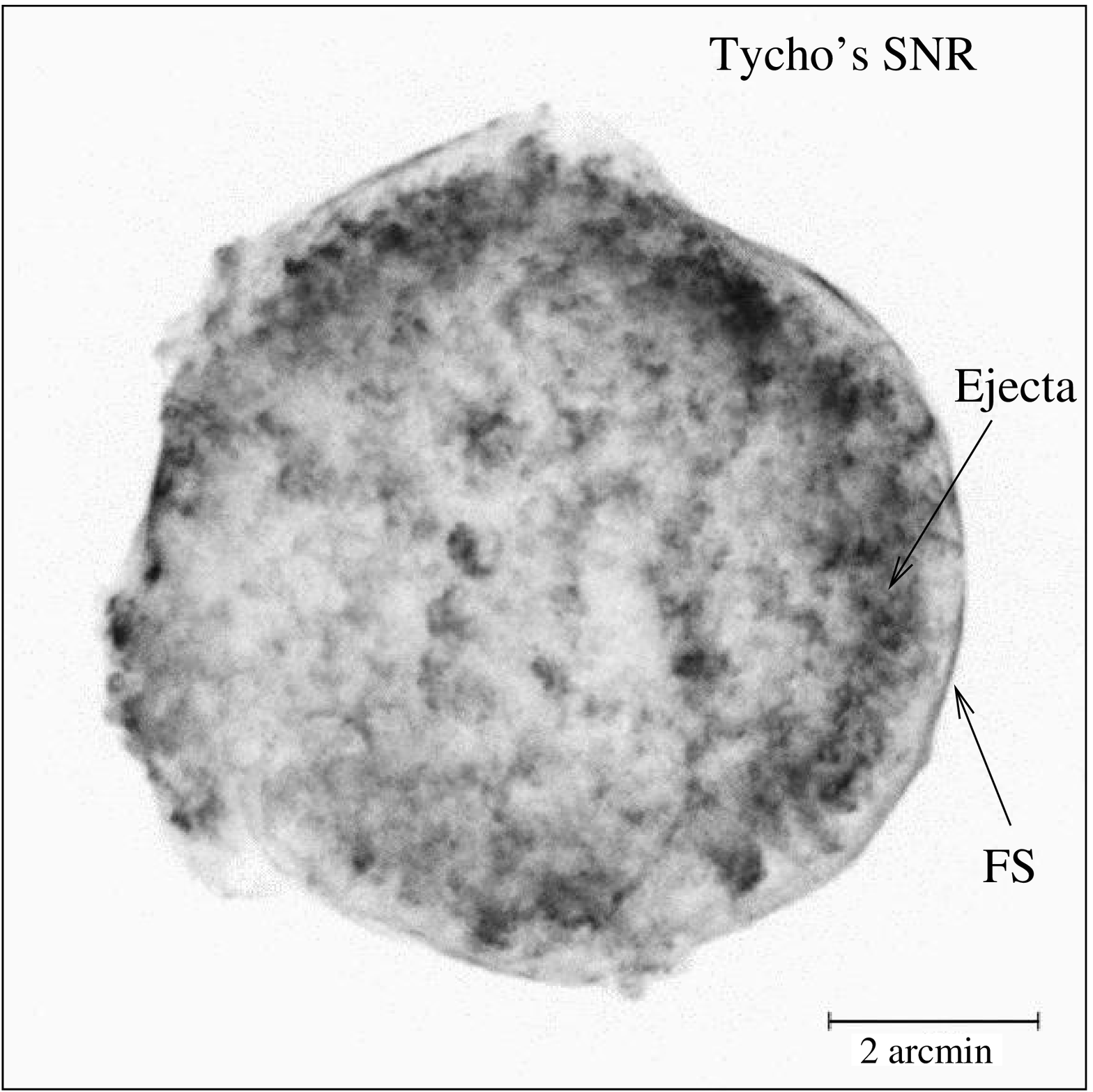}
  }
  \caption{Left: SNR radial temperature distribution for different
values of the particle acceleration efficiency. See text for description.
Right: {\em Chandra} image of Tycho's SNR showing the thin rim of
nonthermal emission at the forward shock (FS), and the inner shell
of shocked ejecta.}
\end{figure}

\section{Nonthermal X-ray Emission from SNRs}

While the shock-heated ejecta and CSM/ISM components of shell-type
SNRs produce line-dominated X-ray spectra, as described above, a
growing number of remnants also reveal evidence of hard, nonthermal
X-ray emission components, apparently associated with high energy
electrons accelerated by the SNR shock.  In several cases -- e.g.,
SN~1006 \cite{Koy95}, G347.3$-$0.5 \cite{Koy97,Sla99}, and G266.2$-$1.2
\cite{Sla01} -- the nonthermal emission components completely
dominate the thermal components, and the X-ray spectra from the
shells are featureless.  For others, including most of the known
young SNRs -- e.g., Cas A \cite{Got01}, Tycho \cite{Hwa02}, and
Kepler \cite{Bam05} -- thin rims of nonthermal emission surround
the remnants directly along their forward shocks. These observations
make it clear that SNRs are capable of accelerating electrons to
very high energies. It is assumed that ions are accelerated by the
same process although, as we describe below, the evidence for this
is less direct. Since ions dominate the cosmic-ray energy density,
however, it is exactly this evidence that is of particular importance
for our understanding of cosmic ray production.

The {\em Chandra} image of Tycho's SNR is shown in Figure 1 (right).
Spectral studies show that the thin outer rim is dominated by
featureless emission consistent with synchrotron radiation from
energetic electrons, while the brighter emission seen in a broad
shell behind the rim is from shock-heated ejecta \cite{War05}.
The position of the reverse shock is along the inner regions
of this ejecta component while the contact discontinuity separating
the ejecta and forward shock is located at the inner edge of the
faint region separating the ejecta and the forward shock.

As discussed above, the separation between the forward shock and
the contact discontinuity is strongly modified by the efficient
acceleration of cosmic ray particles (Figure 1, left). In Tycho's
SNR, this separation is considerably smaller than expected at the
known age of the remnant unless a significant fraction of the
explosion energy has gone into the acceleration of cosmic ray ions
\cite{War05}. These measurements, along with evidence that the
forward shock temperature in the young remnant 1E~0102.2-7219 is
considerably lower than that expected from the measured expansion
velocity \cite{Hug00} (see equation 1 above, which requires $\gamma
\gg 5/3$ for the observed velocity to yield the temperature in this
remnant), provide {\it dynamical} evidence for the acceleration of
cosmic rays ions (which dominate the dynamics because of their large
mass) by SNR shocks; despite the lack of a direct spectral signature
(see below), these observations show that at least some SNRs
accelerate ions, as well as electrons, to multi-TeV energies.

\begin{figure}[t]
  \centerline{
  \includegraphics[height=.3\textheight]{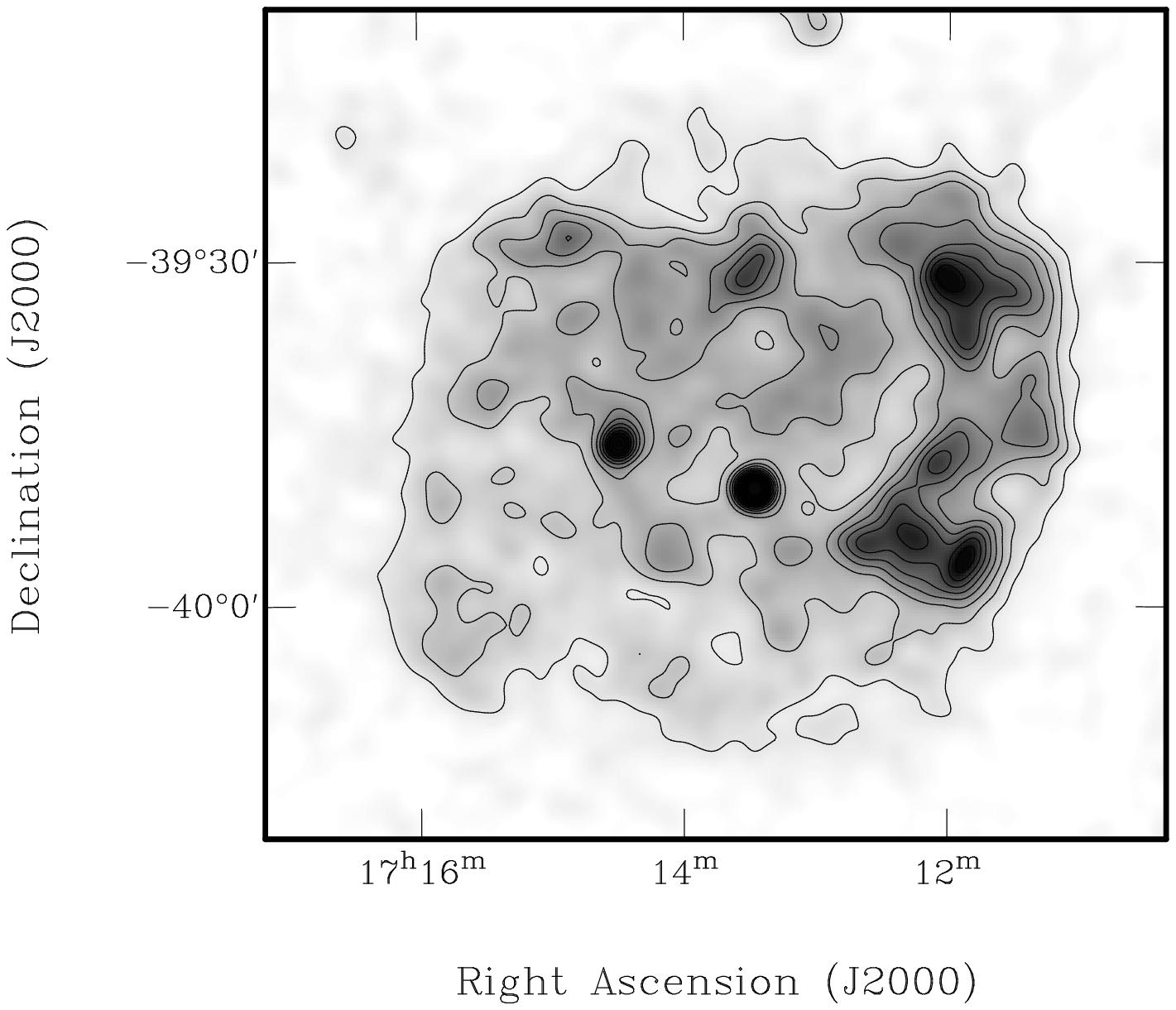}
  \hspace{0.3in}
  \includegraphics[height=.3\textheight]{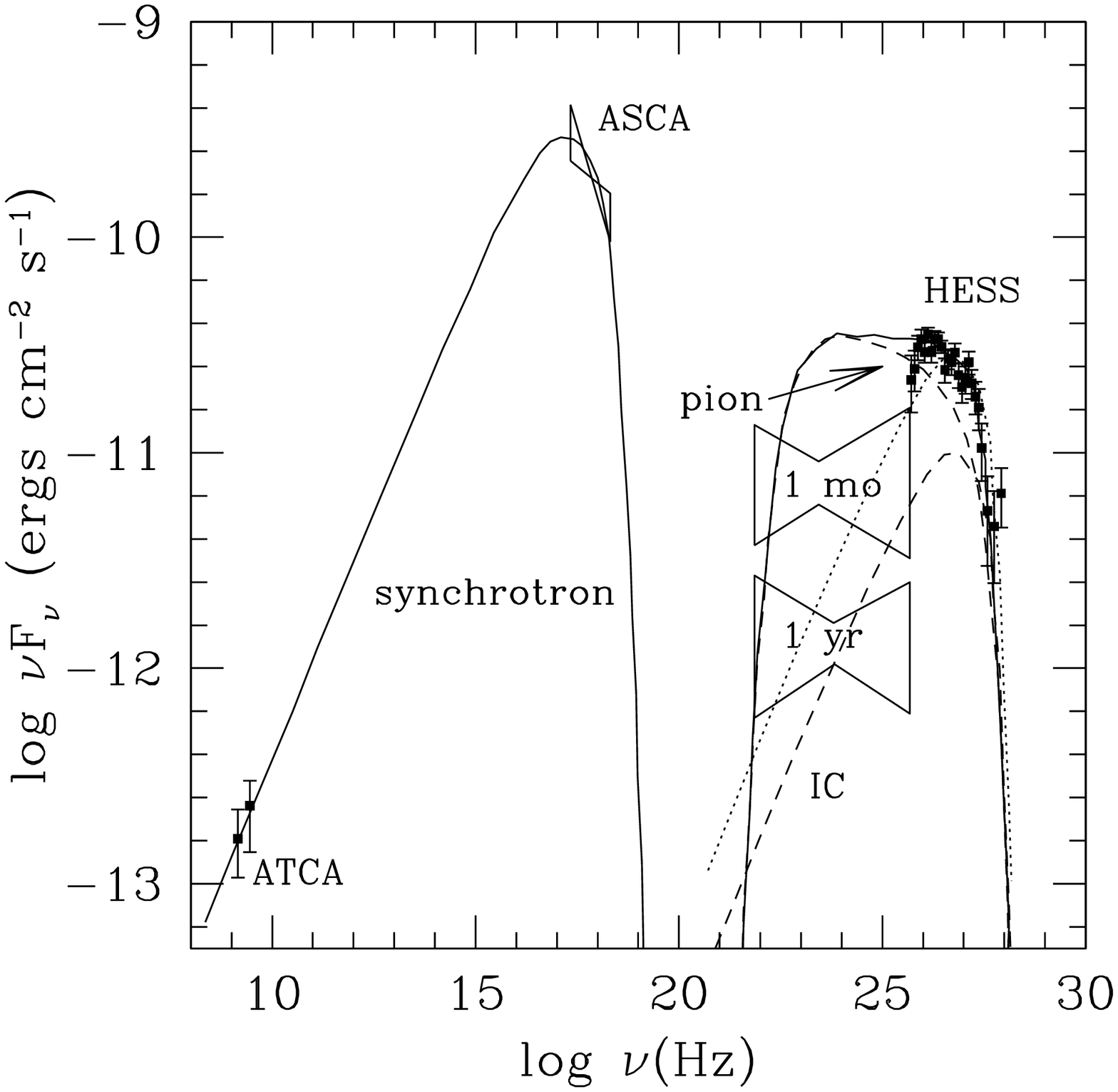}
  }
  \caption{Left: ROSAT PSPC image of G347.3$-$0.5 (RX J1713.7$-$3946). Right:
Broadband spectrum of G347.3$-$0.5. Models for the synchrotron, pion-decay,
and inverse-Compton emission are described in the text. The GLAST sensitivity
is indicated for 1 month and 1 year of all-sky survey data.}
\end{figure}

\section{Gamma-Ray Emission Mechanisms}

At higher particle energies, $\gamma$-ray production may result
from nonthermal bremsstrahlung of electrons colliding with ambient
gas, from IC scattering of ambient photons, and from
the decay of neutral pions created by the collision of energetic
protons. Without additional constraints, it can be difficult to
differentiate between these mechanisms based on $\gamma$-ray spectra
alone. For example, although pion-induced emission leads to a broad
spectrum that, in general, can be distinguished from the more
sharply-peak IC spectrum from a cut-off electron population (see
Figure 2 and the discussion below), high magnetic fields can
produce a steepening of the electron spectrum over a wide energy
range, which can broaden the IC spectrum and make it difficult to
differentiate from a pion-decay spectrum \cite{Ell07}.
The relative contributions from the different $\gamma$-ray production
mechanisms depend highly on ambient conditions, however, and these
conditions also have significant effects on the emission in other
spectral bands. For a strong pion-decay component, for example, a
reasonably high ambient density is required to provide ample target
material with which energetic protons from the SNR shock can collide.
Such dense material should also lead to significant thermal X-ray
emission. A strong IC component, on the other hand, should be
accompanied by synchrotron emission in the radio and X-ray bands
whose luminosity is consistent with expected values for the postshock
magnetic field. In this sense, then, $\gamma$-ray measurements
accompanied by spectra from other bands hold the best promise for
interpreting the high energy emission -- and, in particular, for
providing spectral evidence of high energy ions.

ASCA observations of G347.3$-$0.5 (RX~J1713.7$-$3946) revealed that
the X-ray spectrum of this large-diameter SNR (Figure 2, left) is
dominated by nonthermal emission \cite{Koy97,Sla99}, indicating the
presence of electrons with energies in excess of $\sim 10$~TeV.
Subsequent detection with HESS \cite{Aha06} confirmed the presence
of energetic particles, but the nature of the $\gamma$-ray emission
remains unclear.  The broadband spectrum of the remnant is shown
in Figure 3 (right) along with models for the emission.  The radio
and X-ray emission is well-described by synchrotron radiation from
a power-law distribution of particles accompanied by a high energy
cutoff. The high energy spectrum can be described by a model of
emission induced by both protons (through pion decay)  and electrons
(through IC scattering), shown as dashed curves in Figure 2. Here
the pion component dominates \cite{Mor07}, although the assumed
ambient density ($n_0 = 1 {\rm\ cm}^{-2}$) is hard to reconcile
with upper limits derived from the lack of observed thermal X-ray
emission from the remnant \cite{Sla99}.

A larger overall normalization for the electron spectrum is able
to adequately reproduce the TeV emission from G347.5$-$0.5 without
any pion contribution (dotted curve in Figure 3), but this overpredicts
the synchrotron emission unless a small filling factor is assumed for the
magnetic field \cite{Laz04}. The GLAST sensitivity expected
from all-sky survey mode exposures after 1 month and 1 year are
indicated in Figure 2, and show that the pion component will
either be detected or clearly ruled out within the first year of
observations.

\section{GLAST and SNRs}

\begin{figure}[t]
  \centerline{
  \includegraphics[height=.3\textheight]{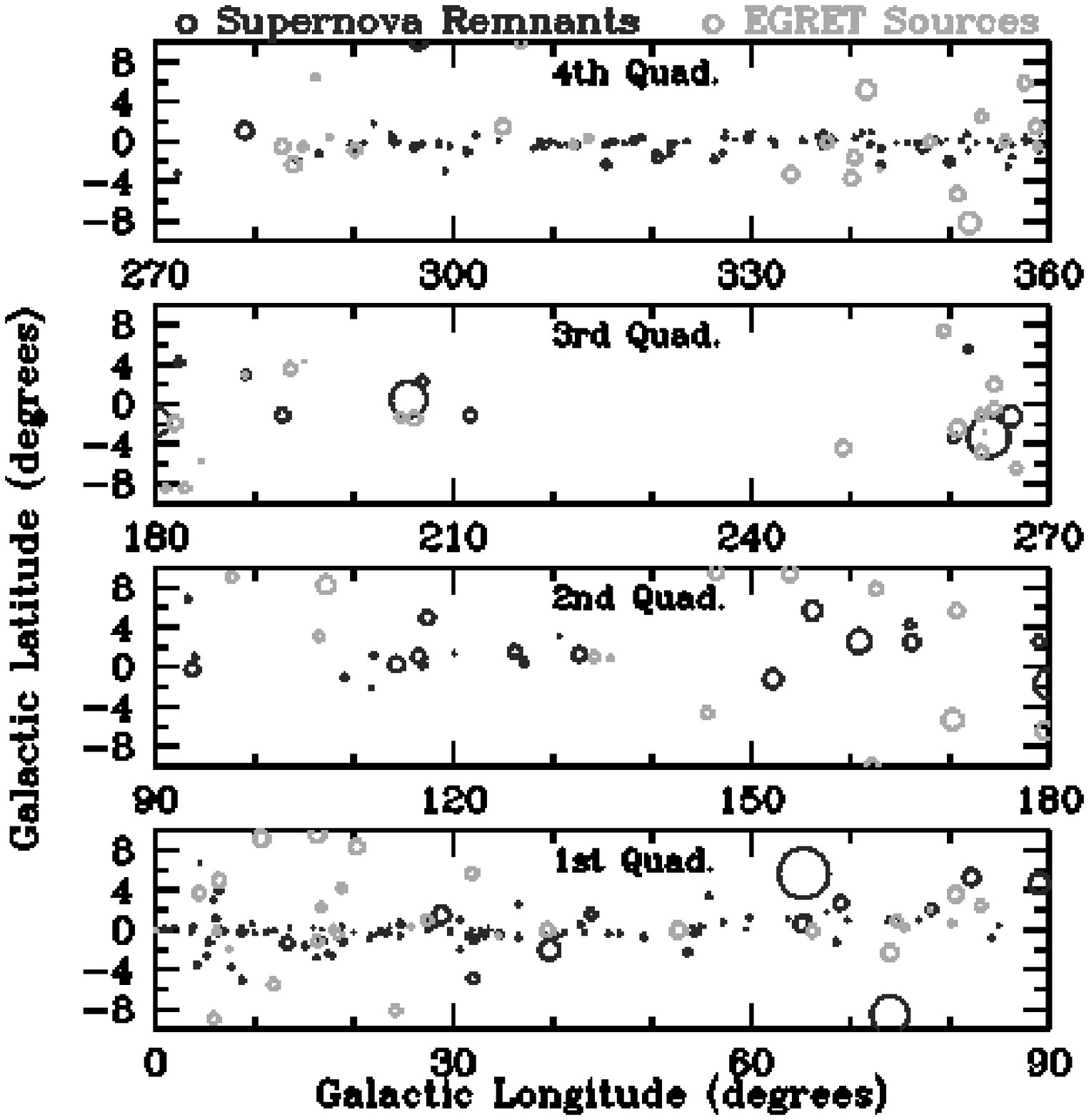}
  \includegraphics[height=.3\textheight]{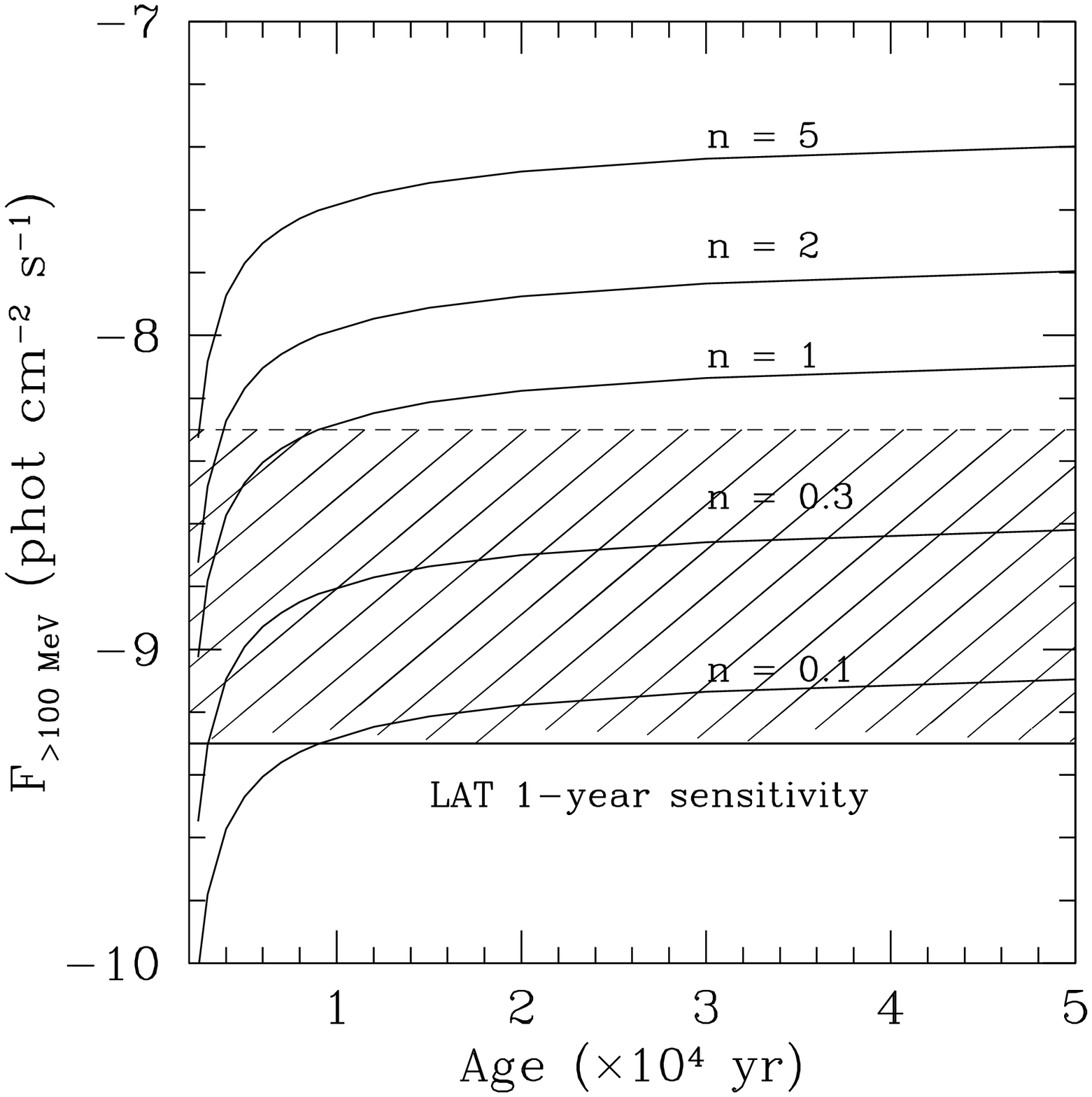}
  }
  \caption{Left: Distribution of supernova remnants (black) and EGRET sources
(grey) in the region of the Galactic plane. Circles indicate SNR sizes and
EGRET source position uncertainties. Right: Predicted pion-decay $\gamma$-ray
flux for SNR expansion into different densities. The solid horizontal line
corresponds to the GLAST survey 1-year sensitivity for a point source. The
sensitivity for extended emission such as that expected from SNRs depends
on the size of the emission region; an arbitrary factor of 10 variation
in sensitivity is illustrated for comparison with predicted fluxes. }
\end{figure}

The distribution of EGRET sources in the Galactic plane has
considerable overlap with that of SNRs (Figure 3, left). Given the
discussion above, this would seem reasonable; we {\it expect} to
detect $\gamma$-rays from SNRs. However, the low angular resolution
provided by EGRET results in large uncertainties for the source
positions. The chance probability that some of these large error
circles would overlap with the position of an SNR is relatively
high, especially since the remnants themselves are extended. Moreover,
the SNR population is closely linked to the distribution of pulsars
and star-forming regions, both classes of which are potential $\gamma$-ray
emitters. 

The expected pion-decay flux from a SNR is approximately
\begin{equation}
F(>100 {\rm MeV}) \approx 4.4  \times 10^7 \theta E_{51} d_{\rm kpc}^{-2}
n_0 {\rm\ phot\ cm}^{-2} {\rm\ s}^{-1}
\end{equation}
where $\theta$ is a slow function of the SNR age, $E_{51}$ is the explosion
energy in units of $10^{51}$~ergs, $d_{\rm kpc}$ is the distance in
kpc, and $n_0$ is the preshock density 
\cite{Dru94}.  For reasonable values of these parameters, the
predicted fluxes are very near the sensitivity limit of EGRET. While
the above estimates are conservative in that any contributions from IC
scattering and bremsstrahlung have been ignored, it remains clear that
only SNRs in particularly dense regions would be expected to produce
detectable emission. Torres et al. \cite{Tor03} have investigated candidate
associations of EGRET sources with SNRs for which there is some evidence
for dense surrounding molecular material, and some of these associations
are indeed plausible. But it is fair to say that there are no unambiguous
detections of SNRs by EGRET.

This somewhat disconcerting view will receive a considerable boost from
GLAST. In Figure 3 (right) we plot the estimated flux in the GLAST band
as a function of SNR age, for a range of ambient densities. The solid
horizontal line represents the LAT sensitivity for detection of a point
source in one year of survey data. The sensitivity is a strong function
of source extent; the hatched region shows a nominal factor of ten 
range in sensitivity to account for different SNR radii. While remnants
in regions with densities typical of the ISM ($n_0 \sim 0.3 {\rm\ cm}^{-3}$),
appear unlikely to be detectable, there is a wide range of densities 
typical of those found in regions of massive star formation that 
should yield detectable SNR emission. For these sources, the prognosis
is quite good: GLAST observations will provide spectra that probe the
important region of the $\gamma$-ray spectrum than can differentiate 
between pion-decay and IC mechanisms, thus making a significant contribution
to our understanding of particle acceleration in SNRs.

\section{Summary}

Several mechanisms exist by which SNRs may be associated with the
emission of energetic $\gamma$-rays. Recent X-ray and TeV $\gamma$-ray
observations have provided powerful evidence for the presence of
high energy particles, and GLAST observations promise contributions
in the crucial spectral range where pion-induced emission can
potentially be discriminated from IC emission. The significant
increase in angular resolution over that provided by EGRET will
provide more confident associations between GLAST sources and SNRs,
and the large collecting area will result in sufficient sensitivity
to accurately determine spectra for comparison with broadband models.
In addition, a significant increase in the discovery space of $\gamma$-ray
emission from SNRs can be expected, allowing us to probe new associations
and potentially resulting in GLAST detections that point the way to
discoveries in other bands.


\begin{theacknowledgments}
I would like to acknowledge the collaborations on this subject that
I have enjoyed with Don Ellison, Dan Patnaude, Steve Reynolds,
Jasmina Lazendic, Jack Hughes, and Bryan Gaensler.
\end{theacknowledgments}


\begin{thebibliography}{9}

\bibitem{Ell07}
D. C. Ellison et al. 2007, ApJ, accepted (astro-ph/0702674).

\bibitem{Koy95}
K. Koyama et al.  1995, Nature 378, 255.

\bibitem{Koy97}
K. Koyama et al. 1997, PASJ, 49, L7.

\bibitem{Sla99}
P. Slane et al. 1999, ApJ, 525, 357.

\bibitem{Sla01}
P. Slane et al. 2001, ApJ, 548, 814.

\bibitem{Got01}
E. V. Gotthelf et al. 2001, ApJ, 552, L39.

\bibitem{Hwa02}
U. Hwang et al. 2002, ApJ, 581, 1101.

\bibitem{Bam05}
A. Bamba et al. 2005, ApJ, 621, 793.

\bibitem{War05}
J. S. Warren et al. 2005, ApJ, 634, 376.

\bibitem{Hug00}
J. P. Hughes, C. E. Rakowski, \& A. Decourchelle 2000, ApJ, 543, L61.

\bibitem{Aha06}
F. Aharonian et al. 2006, A\&A, 449, 223.

\bibitem{Mor07}
K. Moraitis \& A. Mastichiadis 2007, A\&A, 462, 173.

\bibitem{Laz04}
J. S. Lazendic et al. 2004, ApJ, 602, 271.

\bibitem{Dru94}
L. O'C. Drury, F. A. Aharonian, \& H. J. V\"olk  1994, A\&A, 297, 959.

\bibitem{Tor03}
D. F. Torres et al. 2003, PhR, 382, 303.

\end{thebibliography}
\end{document}